\documentstyle[12pt,aasms4]{article}

\received{1998 Aug. 25th}

\def\dfrac#1#2{{\displaystyle\frac{#1}{#2}}}

\makeatletter
\def\lnyoro{\mathrel{\mathpalette\gl@align<}}
\def\gnyoro{\mathrel{\mathpalette\gl@align>}}
\def\gl@align#1#2{\lower.6ex\vbox{\baselineskip\z@skip\lineskip\z@\ialign{$\m@th
#1\hfil##\hfil$\crcr#2\crcr\sim\crcr}}}
\makeatother

\begin{document}

\setcounter{footnote}{-1}

\title{
Photometric Properties of Kiso Ultraviolet-Excess Galaxies 
in the Lynx-Ursa Major Region\footnote{Based on observations 
made at Kiso Observatory, which is operated by Institute of Astronomy, 
Faculty of Science, University of Tokyo, Japan.
}}

\author{
Tsutomu T. Takeuchi$^{1,\, 2}$, Akihiko Tomita$^3$, 
Kouichiro Nakanishi$^1$, Takako T. Ishii$^{1, \,4}$, \\Ikuru Iwata$^5$, and 
Mamoru Sait\={o}$^1$
}

\vspace*{5mm}

\affil{
$^1$ Department of Astronomy, Faculty of Science, Kyoto University,
\\
Sakyo-ku, Kyoto 606-8502, JAPAN
\\
$^2$ Research Fellow of the Japan Society for the Promotion of Science
\\
$^3$ Department of Earth and Astronomical Sciences,
Faculty of Education, \\
Wakayama University,
Wakayama 640-8510, JAPAN
\\
$^4$ Kwasan and Hida Observatories, 
Kyoto University, 
\\
Yamashina-ku, Kyoto 607-8471, JAPAN
\\
$^5$ Toyama Astronomical Observatory, Toyama Science Museum,
\\
49-4 San-no-kuma, Toyama City, Toyama 930-0155, JAPAN
}
\vspace*{5mm}
\affil{
Electronic mail:
takeuchi@kusastro.kyoto-u.ac.jp}

\begin{abstract}
We have performed a systematic study of several regions in the sky where the
number of galaxies exhibiting star formation (SF) activity is greater than 
average. 
We used Kiso ultraviolet-excess galaxies (KUGs) as our SF-enhanced
sample.  By statistically comparing the KUG and non-KUG distributions, we
discovered four KUG-rich regions with a size of $\sim 10^\circ \times
10^\circ$. 
One of these regions corresponds spatially to a filament of length 
$\sim 60\; h^{-1}$ Mpc in the Lynx-Ursa Major region 
($\alpha \sim 9^{\rm h} - 10^{\rm h}\, , \; \delta \sim 42^\circ -
48^\circ$).
We call this ``the Lynx-Ursa Major (LUM) filament''.
We obtained $V(RI)_{\rm C}$ surface photometry of 11 of the KUGs in the LUM
filament and used these to investigate the integrated colors, distribution 
of SF regions, morphologies, and local environments.
We found that these KUGs consist of distorted spiral galaxies and
compact galaxies with blue colors.  
Their star formation occurs in the entire disk, and is not confined to just 
the central regions.
The colors of the SF regions imply that active star formation in the spiral 
galaxies occurred $10^{7 - 8}$ yr ago, while that of the 
compact objects occurred $10^{6-7}$ yr ago.
Though the photometric characteristics of these KUGs are similar to those of 
interacting galaxies or mergers, 
most of these KUGs do not show direct evidence of merger processes.
\end{abstract}


\section{INTRODUCTION}

Kiso Ultraviolet-Excess Galaxies (hereafter KUGs) are a blue galaxy 
sample selected from photographic $UGR$ three-color 
images taken by the Kiso Observatory 105-cm Schmidt telescope  (Takase \&
Miyauchi-Isobe 1993, and references therein, hereafter TM93).
The survey selected galaxies if they had a greater UV excess than A-type 
stars seen on the same plates.
The KUG survey has been carried out mainly on the northern celestial 
hemisphere over $\sim 5800 \; {\rm deg^2}$ , and contains roughly 8200 objects.
The depth of the survey is $17 \sim 18.5$ mag in photographic magnitude.
A second KUG survey is in progress.
A general description of the survey method and the statistical properties of 
KUG objects is given in Takase (1980).

A galaxy that has experienced a star formation (SF) episode less than 
$\sim 10^{8-9}$ yr (sub-Gyr) ago contains an elevated number of A stars and 
so can be identified by a bluer than average color.
The so-called Butcher--Oemler galaxies are mostly in this 
category (Dressler \& Gunn 1983).
Photometric (Maehara et al. 1988), spectroscopic (Maehara et al. 1987, 1988; 
Augarde et al. 1994; Comte et al. 1994), and radio (Maehara et al. 1985, 1988)
observations have revealed that KUGs as a class exhibit star formation. 
Thus, the KUG survey is a good sample of galaxies to use in examining objects 
that have experienced periods of star formation within the recent sub-Gyr in 
the Local Universe.
Tomita et al. (1997)(TTUS97) have quantified the general
characteristics of a large number of KUGs using optical color,
morphology, and FIR data.
Here we briefly summarize their results.
\begin{enumerate}
\item The KUG selection was originally based on a plate search,
and the classification of galaxies into blue and non-blue ones 
proved to work well even in terms of the total color sytems, but 
the boundary color is slightly redder than that of A-stars; 
that is, 22 \% of the KUGs have the non-KUG colors, and
the boundary color separating KUGs and non-KUGs is $(U - V)_{\rm T}
 = 0.1$ mag.
\item KUGs are preferentially Sb- or later-type spiral galaxies.
The KUG fraction changes linearly along the Hubble sequence: 
it is less than 10 \% for E/S0 and more than 50 \% for Sd/Sm.
\item KUGs are biased toward less luminous galaxies.
At around the knee of the luminosity function (LF) where $B$-luminosity $L_B
\sim 10^{10} \,L_\odot$, most of the KUGs are spiral galaxies.
In the fainter regime of $L_B < 10^{9.3}\,L_\odot$, the dwarf population
dominates.
\item The fraction of the blue population in a survey depends on the
depth of the survey.
If the survey is volume-limited and deep enough to pick up the bulk of the 
dwarf population, its fraction would be higher.
\end{enumerate}

The distribution of KUGs is inhomogeneous, not only when compared with a 
uniform distribution but also in comparison with the ambient galaxy 
distribution.  
Consequently there are some ``KUG-rich regions''.
This inhomogeneity may be related to environmental effects.
Some studies show that a dense environment activates star formation
(e.g. Maia et al. 1994; Pastoriza et al. 1994).
On the other hand, recent observational studies (e.g. Zabludoff et
al. 1996(Z96)) suggest that a low-density environment can enhance star 
formation. Therefore, the question is unsettled.

In this study we statistically analyzed the fraction of KUGs in the 
whole galaxy population and discovered four KUG-rich regions.  Among them 
was a region which lies 
in the constellations of Lynx and Ursa Major 
($\alpha \sim 9^{\rm h} - 10^{\rm h}\, , \; \delta \sim 
42^\circ - 48^\circ$).
In this region, there is a galaxy filament with a length of $\sim 60h^{-1}$ 
Mpc (We use $H_0 = 100h \; {\rm km\,s^{-1}\,Mpc^{-1}}$ as the Hubble 
parameter throughout this paper).
We call this structure the Lynx-Ursa Major (hereafter LUM) filament.
We obtained $V(RI)_{\rm C}$ surface photometry of eleven KUGs in this 
region in order to make a quantitative analyses of the star formation 
properties and effects of local environment on star formation.

This paper is organized as follows: Section 2 describes our
statistical survey methods and the KUG-rich regions found. 
Sample selection, observation, and data reduction of the $V(RI)_{\rm C}$ 
photometry of the eleven KUGs in LUM filament are discussed in section 3.
Section 4 presents the results of our photometry.
Discussion based on the photometry is made in section 5.
Finally in section 6, we present a summary of our results.

\section{SURVEY}

\subsection{KUG-rich Regions}

First we searched for regions in which the total galaxy population contains
a large fraction of KUGs.  A single Kiso Schmidt plate has a field of view
of $6^\circ \times 
6^\circ$.  
The previously noted variation in survey depth is caused by differences in 
plate quality.  
To avoid a bias caused by the variation of the photographic-plate depths, 
we required that our sample galaxies also be included in the magnitude-limited 
portion of the {\sl Catalogue of Galaxies and 
of Clusters of Galaxies} (Zwicky et al. 1961 -- 1968, CGCG).
This limited our sample to objects with $m_{\rm pg} \lnyoro 15.7$ mag.
(corresponding to $m_{B_{\rm T}} \lnyoro 15.2$ mag, Kirshner et al. 1978)
at $\delta > -5^\circ$.
Although the faint limit of the CGCG has some uncertainties (e.g. Takamiya 
et al. 1995), they do not affect our analysis.
Hereafter we use the term KUG for galaxies listed in both the CGCG and KUG 
catalogs.

The KUG population fraction, $f_{\rm KUG}$, was defined as 
$f_{\rm KUG} = k/n$ where $n$ = [the total number of CGCG galaxies on a plate]
 and
$k$ = [the number of KUGs on the same plate]. 
We first formed this statistic
for the individual $6^\circ \times 6^\circ$ photographic plates.  
Using only this number, we found some tentative ``KUG-rich'' regions.  
We were still concerned, however, that these could be false effects caused by 
the differing depth of each photographic plate.
To check on this we recalculated $f_{\rm KUG}$ for the same sized areas
in an offset "tessellation" that was
independent of the location of the original Kiso fields.
Both methods yielded almost the same result,
implying that the effects of the plate-quality difference was effectively 
surpressed by using the CGCG subsample. We also formed the $f_{\rm KUG}$
statistic for areas with sizes ranging between $3^\circ \times 3^\circ$ and 
$10^\circ \times 10^\circ$, again with no significant change in results.
Figure 1 shows the distribution of $f_{\rm KUG}$ versus $n$.
Each symbol represents $f_{\rm KUG}$ of a $6^\circ \times 6^\circ$-area 
from the offset tessellation.

The next step was to determine the statistical significance of the 
KUG-rich regions.  As shown in Fig. 1, the mean KUG fraction has a value 
of $\sim 27.7$ \% which is essentially independent of $n$.  
We therefore assumed that $p = 0.277$ represents the mean KUG fraction.
The probability density of $k$ with respect to $n$, $P(n,k)$, follows a
binomial distribution given by
\begin{eqnarray}\label{bino1}
 P(n, k) = \dfrac{n!}{(n - k)!k!} \, p^k (1-p)^{n - k} \qquad \qquad 
\qquad k = 1, \; \cdots 
\; , \; n \; , 
\end{eqnarray}
The standard deviation, $\sigma$, is 
\begin{eqnarray}\label{sig1}
 \sigma = \sqrt{np(1-p)} \quad .
\end{eqnarray}
The probability $P(n, \xi)$ that 
$f_{\rm KUG}$ has the value $f_{\rm KUG} = \xi$ in a population of $n$ CGCG 
galaxies is
\begin{eqnarray}\label{bino2}
 P(n, \xi) = \dfrac{n (n!)}{(n - n\xi)!(n\xi)!} \, p^{n\xi}(1-p)^{n(1-\xi)} 
\qquad \qquad \qquad 0 \leq \xi \leq 1 \; .
\end{eqnarray}
Equations (\ref{sig1}) and (\ref{bino2}) give a standard deviation
for $\xi$, $\sigma_\xi (n)$, of 
\begin{eqnarray}
 \sigma_\xi (n) = \sqrt{\dfrac{p(1 - p)}{n}} \quad .
\end{eqnarray}
The smaller $n$ is, the larger $\sigma_\xi (n)$ becomes.
This is well expressed in Fig. 1 which shows confidence limits 
of 99.9 \%, 99.99 \%, and 99.999 \% as long-dashed, dotted, 
and dashed lines.
We define the KUG-rich regions as those in which $f_{\rm KUG}$ is greater 
than or equal to the 99.99\% confidence limit.
The filled symbols in Fig. 1 represent the KUG-rich regions. 
Eight filled symbols are plotted in Fig. 1, but B overlaps one of the Ds, 
and C overlaps one of the Es making the number of filled symbols appear as six.
Circles with the same labels are regions adjacent to each other on
the sky.
The large extent of the regions and the high confidence limit both 
show that these regions are not mere products of chance. 

An all-sky map containing the KUG-rich regions we discovered is shown 
in Fig. 2.  In this figure
the small dots represent the distribution of CGCG galaxies in
KUG-survey regions, and the black hatches depict the KUG-rich regions.
Their labels (A, B, C, D, and E) correspond to the same lables in Fig. 1. 
KUG-rich region E contains the galaxy cluster Zwicky 1615.8+3505.
Unfortunately, this contains KUG area A0432 (center position: 
$16^{\rm h}20^{\rm m}, \; +35^\circ$), which has been reported to 
contain a sizable fraction of
non-blue galaxies that should not have been in the Kiso survey
(Miyauchi-Isobe, Takase, \& Maehara 1997).
The cluster itself has other interesting aspects, the
details of which will be presented elsewhere (Tomita et al. 1998).
Other four regions have no virialized galaxy structures, i.e., they are in
the ``field''.
Details on region A are presented in section \ref{lumfil}.
Region B seems to be a part of a filamentary structures of galaxies 
in  proximity to Canes Venatici void.
Region C may also belong to a filament surrounding the Gemini void, and
region D may associate with a filament between the Leo and Coma voids that is 
connected to the Coma cluster (Fairall 1998).
We summarize the parameters of these KUG-rich regions in Table 1.
Column 1 gives the labels of KUG-rich regions presented in Fig. 1 and Fig. 2.
Their approximate positions are given in columns 2 and 3, and corresponding 
KUG field numbers are shown in column 4.
Since the offset tessellation defines a grid centered different from the 
original Kiso 
survey field blocks, the black-hatched regions in Fig. 2
and the Kiso fields given in column 4 of Table 1 are at slightly different
positions.

\subsection{The Lynx-Ursa Major (LUM) Filament}\label{lumfil}

Around region A, we found a filament structure of galaxies in 
the CfA survey (e.g. de Lapparent, Geller, \& Huchra 1986).
We name it ``the Lynx-Ursa Major (LUM) filament''
after the constellation it lies in.
This is an elongated and winding structure extending along the line of
sight between $cz \sim 2000 \; {\rm 
km \, s^{-1}}$ and $8000 \; {\rm km \, s^{-1}}$ $(\sim 60 h^{-1} \;
{\rm Mpc})$, at $\alpha \sim 9^{\rm h} - 10^{\rm h}$, $\delta \sim 
42^\circ - 48^\circ$.
The far end of the filament connects to the Great Wall (Geller \& 
Huchra 1989).  
We note that, although ``the Lynx-Ursa Major supercluster'' named by
Giovanelli \& Haynes (1982) and the LUM filament are connected, 
these are actually distinct structures, and the former is much 
larger.
Han et al. (1995) studied the orientation of the spin vectors of
galaxies belonging to a filamentary structure in this area. 
They named it ``the Ursa Major filament''.
Their filament is part of the left leg of 
the ``CfA homunculus'', and differs from our LUM filament.
As mentioned above, region D is associated with Han et al.'s filament.

We summarize the parameters of the LUM filament in Table 2.

\section{OBSERVATION AND DATA REDUCTION}

\subsection{The Sample}

We obtained surface photometry of the KUGs in the LUM filament that met the 
following selection criteria:
(1) an angular diameter $\gnyoro 0^\prime_\cdot 6$, (2) a small inclination 
(${\rm axial \;ratio} \lnyoro 2$).
The first criterion was to ensure the sample galaxies were large
enough to allow the identification and analysis of their star-forming regions.
The second criterion was to minimize the effect of internal
extinction on the objects, the correction of which is uncertain (Buta \& 
Williams 1995, hereafter BW95).  
Eleven objects met these criteria.
We compile their names, positions, and basic properties in Table 3.
Column 1 gives the serial number of the objects, and columns 2 and 3 give 
their KUG and CGCG names respectively.
Columns 4 and 5 give their positions (B1950.0 equinox) extracted from NED.
Columns 6 and 7 give their recession velocity and $B^0_{\rm T}$
Magnitudes, both 
from {\sl Third Reference Catalogue of Bright Galaxies} (de Vaucouleurs et 
al. 1991, hereafter RC3).
The recession velocities are galactocentric.
Column 8 gives the absolute $B$-magnitude of the objects calculated
from $cz$ and $B^0_{\rm T}$.
All of the samples are {\sl IRAS} point sources.

Next we examine whether our selected sample is fair or not.
The LUM filament lies at $cz \sim 2000\; {\rm km\, s^{-1}}$ to $8000
\;{\rm km\, s^{-1}}$.
Among the 11 sample KUGs, No. 1 is the nearest, Nos. 2, 3, 6, 7, 8, and
9 are located in the nearer half of the filament, Nos. 5 and 10 in the
further half, and Nos. 4 and 11 at the further end.
Therefore our sample is not strongly correlated with redshift
in the manner of the KUG and CGCG (see Fig.7 of TTUS97).
Next we check the galaxy luminosities.
The LF of the KUGs show Schechter-like behavior (Schechter 1976), with
a knee at $L_B \sim 10^{10} \, L_\odot$, corresponding to $M_B \sim
-19.45$ mag (TTUS97).
The brightest object is No. 11 ($M_B = -20.71$ mag)
and the faintest is No. 9 ($M_B = -18.20$ mag).
The others distribute rather uniformly between the two.
Our sample consists of slightly bright KUGs, but as a whole, 
it is not strongly biased toward brighter luminosities.
We note that there are no AGNs in the sample.

\subsection{$V(RI)_{\rm C}$ Surface Photometry and Data Reduction}

We obtained $V(RI)_{\rm C}$ surface photometry using the 105-cm Schmidt 
telescope at the Kiso Observatory (hereafter Kiso).
The nights were photometric and had seeing conditions of typically a few 
arcseconds.
The observation log is given in Table 4.
Column 1 shows the names of the sample KUGs.
The observed date is presented in column 2.
Column 3 gives the ID numbers of the obtained orginal CCD image frames at 
Kiso.

The optical system at the prime focus has a focal ratio of $F/3.1$.
We mounted at the prime focus a single-chip CCD camera which uses a TI Japan 
TC215 frontside-illuminated $1000 \times 1018$ chip.
One pixel size corresponded to $0^{\prime \prime}_\cdot 752$, giving a total 
field of about $12^\prime_\cdot 5 \times 12^\prime_\cdot 7$.
Nonlinearity was $< 0.5$ \% at $< 25000$ counts, which is too small to give 
rise 
to any significant photometric error. 
We used broadband Johnson $V$, Cousins $R_{\rm C}$, and $I_{\rm C}$ 
filters.
The Kiso $V$-, $R_{\rm C}$-, and $I_{\rm C}$-band response functions are in 
good 
agreement with the standard filter transmissions so we did not need any color
corrections.
The exposure time for all objects and filters was 900 sec. 

The basic data reduction consisted of a bias subtraction, flatfielding, 
and cosmic-ray elimination, using IRAF\footnote{IRAF is distributed by 
National Optical Astronomy Observatories (NOAO), which is operated by the 
Association of Universities for Research in Astronomy, Inc., under 
contract to the National Science Foundation.}.
Sky subtraction was accomplished via a polynomial-surface fit to the 
sky regions using SPIRAL (Hamabe \& Ichikawa 1992).
We made various tests to estimate the uncertainty of the sky subtraction, 
which turned out to be $\lnyoro 1$ \%.

Flux calibration was done by observing the equatorial photometric 
standard stars of Landolt (1992) throughout the night.
The differences of the standard magnitudes from the original 
values of Cousins for our $V$-, $R$-, and $I$-bands are
small enough (Menzies et al. 1991), to preclude making any 
corrections.
We obtained integrated magnitudes for each object from the asymptotic 
limit of a circular-aperture curve of growth after any stars were removed 
from the galaxy image. 
We used the template curve of growth provided by Kodaira, Okamura, \& Ichikawa
(1990) for testing the convergent value of the curve of growth.
The reliability of this method is examined in Tomita et al. (1998).
The total error induced by all procedures is typically $\sim 0.03$ mag.

In order to better estimate the colors of the objects, we corrected 
for differences caused by variations in the seeing size and position between 
the images taken in different bandpasses.
We used IRAF and SPIRAL for this procedure.
For the detailed color analysis we used IDL (Interactive Data Language).

\section{RESULTS}
\subsection{Total Magnitudes by Growth-curve Fitting}

We show the results of the photometry in Table 5.
Column 1 gives the name of the galaxy. 
Columns 2, 3, and 4 give the apparent integrated $V$-, $R_{\rm C}$-, 
and $I_{\rm C}$-magnitudes respectively.
No correction for internal or external extinction has been applied for the 
{\sl apparent} magnitudes.
Columns 5, 6, and 7 give the derived total absolute magnitudes after 
correction for Galactic extinction. The Galactic extinction value at Lynx -- 
Ursa Major, derived from the maps of Burstein \& Heiles (1982),
is $E(B - V) \sim 0.03$.
This corresponds to $A_V \sim 0.09$ mag, $A_R \sim 0.07$ mag, and $A_I \sim 
0.04$ mag when we apply $R_V = A_V / E( B - V) = 3.1$, using the extinction 
curve of Cardelli, Clayton, \& Mathis (1989).
Absolute magnitudes are dereddened by these values.
We did not apply any correction for inclination.
No attempt has been made to apply $K$-correction because of the closeness 
of our sample.

These are the first photometric measurements for all but one of these objects,
KUG0953+466.
This object is also Mrk 129 (Markarian et al. 1989), 
and has Johnson $V$- and $R$- band photometry from Huchra (1977)(H77).
Using 24$''$-aperture photometry, the same aperture size as H77, 
we measured the $V$- and $R_{\rm C}$-magnitudes for comparison.
Then we converted his $R$-magnitude to $R_{\rm C}$.
(We also use this conversion in section 5.2, where a detailed discussion 
of the conversion formula will be given.) 
Table 6 shows our results and shows that our measurements are  
identical with the values from H77, to within the errors.

\subsection{Contour Maps}

Figure 3 shows contour maps in the $V$- and $I_{\rm C}$-bands of all 11
objects in our sample.
They are shown from top to bottom in increasing order of right 
ascension, $\alpha$.
The contour interval is 0.5 magnitudes.
In this figure, north is above, and east is to the right.
The FOV is $2^\prime_\cdot5 \times 2^\prime_\cdot5$.
A distance scale bar of $10\,h^{-1}$ kpc is presented in the upper-left 
of each galaxy image.
We comment on the morphological feature of each galaxy.  
Unless otherwise noted, the morphological index is the $T$-type from the RC3. 

\paragraph{KUG 0908+451 (NGC 2766, IRAS F09089+4509)} 

The morphological index is $T = 5$.
This galaxy is quite knotty, with a large number of H{\sc ii} regions.
This object has a globally distorted appearance, and well-developed 
arms. The eastern arm is extremely blue.
There are no companions in the FOV of the Kiso CCD (corresponding to 
$\sim 100 h^{-1} \; {\rm kpc} \times 100h^{-1} \; {\rm kpc}$).

\paragraph{KUG 0908+468 (Mrk102, IRAS F09082+4650)}

A morphological index is not given in the RC3, and
cannot be accurately estimated.
This is a featureless, spheroidal galaxy, with a slightly boxy isophote.
It is isolated in $\sim 160h^{-1} \; {\rm kpc} \times 160h^{-1} \; {\rm kpc}$.

\paragraph{KUG 0911+471 (UGC 4870, IRAS F09115+4706)}

A morphological index is not given in the RC3.  
We estimate a value of $T \sim 5$.  
The disk is slightly warped.
This galaxy has H{\sc ii}-region knots in its 
peripheral region around the disk which are not well presented in the
smoothed contours.
There are no galaxy companions in $\sim 160h^{-1} \; {\rm kpc} \times 
160h^{-1} \; {\rm kpc}$.

\paragraph{KUG 0919+474 (Mrk109, IRAS F09190+4727)}

A morphological index is not given in the RC3, and
cannot be accurately estimated.
This object has a compact and featureless appearance.
It lies in a loose group of galaxies and has some companions.

\paragraph{KUG 0924+448 (UGC 5045, IRAS F09249+4452)}

The morphological index is $T = 5$.
This object is a triple-arm barred galaxy which suffers from global
distortion.
Its arms are knotty with many H{\sc ii} regions.
It is isolated in a field of $\sim 270h^{-1} \; {\rm kpc} \times 
270h^{-1} \; {\rm kpc}$.

\paragraph{KUG 0944+468 (UGC 5237, IRAS F09441+4650)}

The morphological index is $T = 6$. The object is asymmetric.
The arm in the south is well-developed with large blue knots.
It is isolated in a field of $\sim 160h^{-1} \; {\rm kpc} \times 
160h^{-1} \; {\rm kpc}$.

\paragraph{KUG 0945+443 (NGC 2998, IRAS F09455+4418)}

The morphological index is $T = 5$.
This galaxy has extremely well-developed knotty arms.
It lies in a loose group, and there is a faint irregular galaxy 
lying at $5'$ ($\sim 60h^{-1}$ kpc) to the east.
They do not appear to be bridged, or have any features which imply 
interaction.

\paragraph{KUG 0947+445A (NGC 3009, IRAS F09470+4431)}

A morphological index is not given in the RC3.  
We estimate a value of $T \sim 4$. 
This galaxy has four-arms, with the one in the north being less 
developed than the others.
The appearance is rather smooth with no prominent H{\sc ii}-regions.
It lies in a loose group, and has small companions in almost the same 
redshifts, but without any tidal features.

\paragraph{KUG 0953+466 (Mrk129, IRAS F09535+4641)}

A morphological index is not given in the RC3, and
cannot be accurately estimated.
This object is a featureless, boxy-shaped compact galaxy.
It is isolated in $\sim 160h^{-1} \; {\rm kpc} \times 
160h^{-1} \; {\rm kpc}$.

\paragraph{KUG 1007+461 (NGC 3135, IRAS F10078+4611)}

A morphological index is not given in the RC3.  We estimate a value of 
$T \sim 5$. Its arms are ill-developed.
This galaxy has a companion in the north-east.

\paragraph{KUG 1016+467 (NGC 3191, IRAS F10159+4642)}

The morphological index $T = 4$.
This galaxy suffers from heavy distortion, and has a companion to the west.
There is an extremely blue tidal bridge between them suggesting that 
they are interacting.

\subsection{Colors}

We give the total integrated $V(RI)_{\rm C}$ colors of the objects in Table 7.
They are corrected only for Galactic extinction.

We then further divided each image into sections 
and derived colors section-by-section in order to investigate 
the distribution of SF regions within each galaxy.
An SF region that is $2-3$ kpc diameter (which is typical of such regions) 
has an angular size of $\sim
7''$ at $cz \sim 9000 \; {\rm km\,s^{-1}}$, 
the distance of the furthest galaxy in our sample.
This angular size is much larger than the seeing disk of our images, 
therefore we can divide the most distant image into $7'' \times 7''$-sized
squares without oversampling.  For consistency we scaled the sample 
resolution applied to our nearer objects with redshift to match the same 
physical size of 3 kpc.
Then, for the regions which have $S/N > 3$, we evaluated the colors.
Figure 4 shows the partitioning of each galaxy.
Color-color (C-C) diagrams of the sections in each galaxy
are presented in Fig. 5. 
The dominant error of the color estimation comes from the sky fitting.
Points labeled ``bulge'' and ``disk'' are from those sections of each
galaxy respectively.
For some representative regions in each galaxy, we have labeled its number
from Fig. 4 beside the corresponding point in Fig.5.
The three broken lines on the C-C diagrams show starburst evolutionary
tracks superimposed on an old stellar population (Bica,
Alloin, \& Schmidt 1990, hereafter BAS90).
Detailed discussion about these tracks will be given in section 5.2.

\section{DISCUSSION}

\subsection{Total Colors}

We compare the total colors of our sample with those of galaxies
listed in the RC3.
We have taken our values for $(V - R)^0_{\rm T}$ and $(V - I)^0_{\rm T}$
From BW95.
Their filters are the same as those of Kiso, Johnson $V$, and 
Cousins $R_{\rm C}$ and $I_{\rm C}$, so it was not necessary to apply
a correction.  However, they corrected the 
galaxy colors for internal extinction and inclination.
To compare our results with the data from by BW95, we 
applied a correction of $A_V \sim 0.15$ for internal extinction.

Fig. 6 plots $(V - R)^0_{\rm T}$ versus $(V - I)^0_{\rm T}$ for 
the KUGs in the LUM filament together with
those of a wider range in Hubble type from BW95.
Open circles represent the mean color of each morphological type index $T$
\footnote{The shown indices are $-5$, $-4$, $-3$, $-2$, $-1$, 0, 1, 2, 3, 4, 
5, 6, \{7 and 8\}, 9, and 10, according to BW95.}.
The errors bar give the standard deviation in each color across the
galaxies belonging to each morphological type.
Filled squares represent the dereddened colors of our KUGs.
The applied reddening correction vector is presented in the top left of 
the diagram.  
The typical photometric error of our data is shown in the bottom right. 
The reddening correction value is set to compare our photomety with that of
BW95.
The morphological $T$-index of each KUG is indicated.
For the six objects without $T$ in RC3, we use our own classification
or label them as ``C'' for compact morphology if we could not determine a
$T$ value.  
The colors of our sample tend to deviate systematically blueward from the
range of the $1\sigma$-strip of the Hubble sequence in $(V - I)_{\rm T}$.
Even though our sample consists of intermediate/late-type spiral and 
compact galaxies, they are scattered well outside the distribution of 
the BW95 sample.
This is attributed not only to the blue continuum emission from 
early-type stars but also to the strong emission lines from star forming
regions.
The strong H$\alpha$ line makes their $(V - R)$ color redder, which is 
clearly seen in Fig. 6.

A similar tendency has been reported for other UV-excess 
galaxy samples.
Markarian galaxies are dispersed on a $(U - B)$-$(B - V)$ diagram wider
than field galaxies (H77).
Barth, Coziol, \& Demers (1995, hereafter BCD95) have mentioned that their 
Montreal Blue Galaxy (MBG) sample shows a larger dispersion than normal
galaxies on the $(B - V)$-$(V - R)$, and $(B - R)$-$(V - I)$ planes.
The MBG is a blue galaxy sample selected by a similar but more restricted 
method than the KUG survey(Coziol et al. 1996, and references therein).
Larson \& Tinsley (1978) pointed out that interacting galaxies show a large 
scatter on C-C diagrams.
Therefore, BCD95 have suggested a link between active star formation in MBG 
and galaxy-galaxy encounter, since such behavior of interacting galaxies is 
similar to those of their samples.
Our analysis, however, shows that the blue color of KUGs is not neccesarily
connected to interactions (see section \ref{mor}).

\subsection{Distribution of SF Regions}\label{distsf}

Figure 5 clearly shows that all the spiral KUGs of our
sample have bulges redder than their disks, like normal galaxies.
This means that the LUM-filament KUGs are not starburst nucleus 
galaxies (SBNGs), a fact consistent with the KUG 
morphological types given in the original catalog.  
This is in contrast to the fact that the most MBGs are SBNGs 
(Coziol et al. 1996).  A detailed discussion on morphology is 
in section \ref{mor}.
The rest of our sample turns out to be compact galaxies.
These compact galaxies are not too distant to judge their morphology
($cz \sim 5000 \; {\rm km\, s^{-1}}$), and are in reality
featureless.
Their properties are similar to those of the so-called blue compact
dwarf galaxies (BCDGs), but our compact KUGs have $\log L_B > 9.5$
($M_B < -18.2$ mag), brighter than BCDGs.
This is consistent with the global properties of the whole KUG survey 
(TTUS97, and references therein).

A comparison of the starburst evolution models with the color distribution
in Fig. 5 provides an estimate of the onset of star formation in each object.
As previously noted, we used evolution tracks from BAS90, which 
describe the spectral evolution over $3 \times 10^9$ yr of a starburst in low 
metallicity gas, superimposed on an older stellar population. 
Their models combine elements from star cluster and 
galaxy spectral libraries.
A star cluster of a given age defined the starburst spectral signature
while a red galaxy nucleus represents a typical old metal-rich underlying
population.
The flux proportions for combining the spectra are dictated by three 
burst-to-galaxy mass ratios of 10\%, 1\%, and 0.1 \%.
In Fig. 5, solid, short-dashed, and long-dashed lines represent 10-\%, 1-\%, 
and 0.1-\% burst masses, respectively.

BAS90 calculated Johnson $BVRI$ colors from their synthesized 
spectra.  We converted their starburst colors into the Cousins system via
the following two formulae:
\begin{eqnarray}
 (V - R)_{\rm JC} &=& \frac{1}{1.40} \, 
\left( (V - R)_{\rm J} - 0.028 \right) \; ,
\quad (V - R)_{\rm JC} < 1.0 \\
 (V - I)_{\rm JC} &=& \frac{1}{1.30} \, 
\left( (V - I)_{\rm J} - 0.013 \right) \; ,
\quad (V - I)_{\rm JC} < 2.0
\end{eqnarray}
(Cousins 1976), and 
\begin{eqnarray}
 (V - R)_{\rm JC} &=& 0.73 (V - R) _{\rm J} - 0.03 \; , 
\quad (V - R)_{\rm J} < 1.0 \\
 (V - I)_{\rm JC} &=& 0.778 (V - I) _{\rm J} - 0.03 \; , 
\quad (V - I)_{\rm J} < 2.0 
\end{eqnarray}
(Bessell 1979).
The converted colors derived from these two formulae agree
within $0.01$ mag.
The tracks on Fig. 5 are the converted evolutionary loci.
Furthermore, we need to estimate the internal extinction for comparison.
As with Fig. 6, an applied reddening vector of $A_V = 0.1$ is shown in 
the upper left corner.

Figure 5 shows that most of the partial colors of our sample KUGs are
located on the BAS90 model tracks, affirming the validity of applying
this model to out measurements.
We emphasize that the models are of a starburst superimposed
on an old population, instead of the evolutionary models of 
the starburst component alone.
Though we should keep in mind that the galaxy color is sensitive to the 
burst-mass fraction (BAS90) and metallicity of the gas (Leitherer \& Heckman 
1995),
Both of which are rather uncertain, the important fact is that most of
our sample KUGs have a young ($\lnyoro 10^{7-8}$ yr) stellar component
within their {\sl old disks}, i.e. the star formation is taking
place in the disks of the KUGs, not in their central regions.

We next examine the color distribution of the individual KUGs.

\paragraph{KUG 0908+451}

This is a giant grand-designed spiral galaxy.
The partial colors of the disk distribute around the $10^8$-yr phase of 
the 10-\% burst mass track, or $10^7$-yr phase of the 1-\% burst mass one.
The symbol labeled 12 corresponds to the bulge of this galaxy.
Clearly, the color of the bulge is red, implying a lack of a young population.
The arm of the galaxy (labelled 23 in Fig. 5) is abundant in H$\,${\sc ii} 
regions, and turns out to be the very blue color, corresponding to 
a stellar age of less than $10^7$ yr of a 10-\% burst mass.

\paragraph{KUG 0908+468}

This galaxy is compact with the same color throughout.
Comparison with the tracks indicates that the stellar ages of the 
burst component are $<10^8$ yr, and burst mass ratio
to the total baryonic mass is high ( $> {\rm a\; few}$ \%), i.e. ongoing
starburst occurs over the whole galaxy.

\paragraph{KUG 0911+471}

This galaxy belongs to the giant spirals.
The disk color is located around the $10^8$-yr phase of the 
10-\% burst mass track.
The symbol labeled 6 represents the color of the bulge, 
which has a normal color for an old stellar population.

\paragraph{KUG 0919+474}

This galaxy is compact, and like KUG 0908+468, has the same color througout. 
The color of this galaxy can only be explained with a very high 
burst mass ratio, $\sim 10$ \%.
The implicated burst time is very recent ($10^6 \sim 10^7$ yr).

\paragraph{KUG 0924+448}

This is a giant spiral galaxy with a variety of colors in the disk.
Most of the disk color clusters around the $10^8$-yr phase of the 
$1 \sim 10$ \% burst mass tracks.  Some other parts show colors 
corresponding to younger stellar populations.
As with other giant spiral KUG samples, the bulge of this galaxy
(labeled 20) is red.

\paragraph{KUG 0944+468}

This is also a giant spiral galaxy, though it looks small in size because of 
its rather low surface brightness.  Its bulge (labeled 4) is bluer than comparable KUG samples.  But its disk parts are still bluer than the bulge.
The color corresponds to a burst more recent than $10^8$-yr on the 
$1 \sim 10$-\% tracks.

\paragraph{KUG 0945+443}

This is also a giant spiral galaxy.
The dispersion of colors across the galaxy is large, implying that the
ages of the SF regions on the disk range from $5 \times 10^7$ yr to $10^9$ 
yr.
The most crowded domain on the C-C diagram corresponds to the $10^8$-yr phase 
of the 10-\% or 1-\% burst mass tracks.
Its bulge (labeled 12) is red, as expected from an old stellar component.

\paragraph{KUG 0947+445A}

This galaxy is a spiral, but has a rather 
strange shape consisting of four separate arms or two 
arms with a bar crossing between them.
In addition, it is as small as a compact galaxy.
Despite its morphological peculiarity, its disk parts have rather old 
stellar colors, corresponding to ${\rm several} \times 10^8$-yr phase of 
the $1 \sim 10$-\% burst mass tracks.
Its bulge (labeled 5) has the same color as its disk parts, though 
the color is not so blue that it would be regarded as a nuclear starburst.

\paragraph{KUG 0953+466}

This is a compact galaxy with a large dispersion in $(V - I)_{\rm T}$. 
Its bursts correspond to a stellar age of $\sim 10^{6 - 7}$ yr with 
a burst mass fraction of ${\rm several} \times 0.1$ \%.
The partial colors are well explained by this burst age and mass fraction,
therefore the age dispersion may be also small as those of other compact 
KUG samples.

\paragraph{KUG 1007+461}

This is a giant spiral galaxy.
Its bulge (labeled 9) is slightly bluer than those of other spiral galaxy 
samples, and its disk components distribute
around the domain of a very young stellar 
population.
Suggested stellar age is around $10^7 \sim 10^8$ yr.

\paragraph{KUG 1016+467}

This is the only interacting galaxy pair in our sample.
The parent or larger one is a giant spiral galaxy.
We show only the partial colors of the parent galaxy in Fig. 5. 
This galaxy has a large color dispersion across its disk, corresponding to 
a stellar population with an age of $10^7 \sim {\rm several} \times 
10^8$ yr superimposed on an old background population.
Its bulge (labeled 13) is also somewhat blue.

\subsection{Morphology}\label{mor}

We have described the individual morphologies of the LUM-filament KUGs
in section 4.2.
In addition, the KUG catalog gives original morphological 
classification as follows (TM93):
Ic is an irregular galaxy with clumpy H{\sc ii} regions, Ig is an irregular
galaxy with a conspicuously giant H{\sc ii} region, Pi is a pair of 
interacting components, Pd is a pair of detached components, Sk is 
a spiral galaxy with knots of H{\sc ii} regions along its arms, Sp is 
a spiral galaxy with a peculiar bar and/or nucleus, C is a compact galaxy, 
and ``?'' is unclassifiable.
The morphological properties of the objects in this study, 
as well as the originally cataloged KUG morphology and star formation properties 
are summarized in Table 8.
Column 1 shows the KUG names.
Column 2 presents the RC3 $T$ indices for each KUG.
For the galaxies which RC3 $T$s are not given, we give our assigned
indices instead.
Unclassifiable compact galaxies are labeled ``C''.
We show the KUG morphological classification in column 3.
The star formation properties are presented in columns 4 $-$ 6;
column 4 gives the location where star formation
occurs within the galaxy, column 5 gives
the derived ages of the superposing stellar population, and column 6 gives 
the burst-to-galaxy mass ratios of our samples.
We note their morphological features in columns 7, 8, and 9.
Our LUM-filament KUG sample are roughly divided into two classes, which are
the class of knotty spirals Sk and the class of compact galaxies C.
This is consistent with the morphological analysis based on our CCD images.

Despite the variety of morphologies, some general points are apparent:

\begin{enumerate}
\item Our sample KUGs are divided into two classes, giant spiral systems 
and compact galaxies.
\item Most of the giant-spiral KUGs show disk distortion.
\item Giant-spiral KUGs have highly developed arms and a
knotty appearance due to giant H{\sc ii} regions.
\item In spite of 2 and 3, roughly half of our sample is isolated, 
and the other half has some companions, or lies in a group of galaxies.
\item Only one of our sample (KUG1016+467) shows explicit interaction 
features.
This galaxy is an interacting system of a giant spiral and a small amorphous 
satellite.
\end{enumerate}
One of the main conclusions of TTUS97 is that, in terms of stellar
population, the late-type KUGs are mostly normal galaxies and the
early-type KUGs often turned out to be peculiar galaxies.
The color difference between the KUGs and non-KUGs is significant for
$T<5$, and indicates that the early-type KUGs have a young stellar population
for their morphologies.
Compact galaxies tend to be included in the early-type galaxy classification
as peculiar early-types and are thus not really a proper member of the class.
On the other hand, though the spiral KUGs in our sample have a young
stellar population in their disks which makes their appearance quite
knotty, they are rather normal for their morphology.
This is also consistent with the above result of TTUS97.
The spiral KUG samples have no active nuclei or circumnuclear
Starbursts. Therefore they are ``active normal spirals''.

\subsection{Correlation between Various Properties of our Samples}

Here we examine the relation between the colors, ages, morphologies, 
and environments of our samples.
Many authors have claimed a close connection between active star formation and 
morphological distortion.
For example, BCD95 have performed a quantitative morphological 
analysis of their MBG starbursts by Fourier transforming the isophotes of 
sample galaxies.  They concluded that the star formation regions 
associate with isophotal twists.
Such distortion or twist from symmetry is also found in our KUG sample in 
the LUM filament.
Z96 have discovered that a significant fraction of their ``E+A'' 
sample exhibits tidal features.
But, as BCD95 has pointed out, the morphological distortion is not a 
direct evidence of galaxy-galaxy interaction.
Half of our sample is isolated galaxies.  Only one has a
tidally-connected companion.
Thus, we cannot attribute the starformation in our sample to distortion features, though galaxy
encounters might be responsible for some of the starbursts.

We next consider the relations between the stellar ages implied by the 
color distribution and other properties.
A clear morphological relation exists. Compact galaxies 
which have a featureless appearance have very young superposing stellar 
components ($10^6 \sim 10^7$ yr), i.e. on-going starbursts, spread over 
the whole galaxy.
In contrast, giant spiral galaxies which have well-developed arms have 
more aged stellar components ($10^7 \sim 10^8$ yr) in their disks, 
and their bulge is old.
The age scatter is large in each spiral sample.
But there is an exception, KUG0947+445A, which is a spiral galaxy but has 
a small dispersion of partial colors.
We note that this is a physically small galaxy, as we saw in section 
\ref{distsf}.
The color dispersion appears to correlate with galaxy size with larger galaxies
having a greater dispersion. 

Finally we look into the relation between the stellar age and existence of 
companions.
The most active star forming galaxy is KUG 0919+474, which is a compact galaxy in a group having a burst age of $10^6 \sim 10^7$ yr.
Isolated compact galaxy KUG 0908+468 also has rather young starbursts, 
but their ages are not as young as those of KUG 0919+474.
Another isolated compact KUG 0953+466 has a burst age of 
$10^6 \sim 10^7$ yr, with a small burst mass fraction of ($\sim {\rm several} 
\times 0.1$ \%) implied.
Spiral galaxies have similar stellar ages with each other.
Some KUGs in a group or with companions have rather younger stellar ages, 
like KUG 1007+461 and KUG1016+467.
In contrast, other KUGs with companions have the superposing stellar 
components consistent with intermidiate ages, like KUG 0945+443 and 
KUG 0947+445A.
Thus, there is no clear relation between the superposing stellar age 
and the existence of companions.

\section{SUMMARY}

We have focused on the collective star formation enhancement of galaxies, and
searched for the regions where such phenomena have occurred using the Kiso
Ultraviolet-excess Galaxy (KUG) catalog as a SF-enhanced galaxy
sample.
Through our survey we found four KUG-rich regions, one of which
turned out to be
associated with a filamentary structure.
We named it ``the Lynx-Ursa Major (LUM) filament'' after its location on the 
sky ($\alpha \sim 9^{\rm h} - 10^{\rm h}\, , 
\; \delta \sim 42^\circ - 48^\circ$).
We then investigated the star formation properties of the KUGs in the LUM.
Our results are as follows:
\begin{enumerate}
\item The eleven KUG objects we examined in the LUM filament proved to be
generally blue by CCD photometry, 
and they show a much larger scatter on C-C diagrams than that of normal 
galaxies.
This is similar to those of other UV-excess galaxy surveys, such as the 
Markarian survey, or MBG survey.

\item The spiral subset of the sample has conspicuous H{\sc ii}-region knots 
in their arms.
The rest are compact galaxies with no structure and an extremely 
blue color.
We suggest that the former corresponds to the KUG subset of ``normal 
late-type spiral galaxies'', and the latter corresponds to the 
``peculiar early-type galaxies'', proposed by TTUS97.

\item The star formation of our sample KUGs occurs in the whole disk, 
and is not concentrated in the central regions.
This is different from those of SBNGs which are the main constituent of 
The MBG survey.
None of our sample KUGs have nuclear activity.
The age of their young stellar population is $\lnyoro 10^{7-8}$ yr.
Compact samples have very young superposing stellar components 
($10^6 \sim 10^7$ yr) spread over the whole galaxy, while
giant spiral samples have moderately aged stellar components 
($10^7 \sim 10^8$ yr) in their disks, and and an older bulge.
The age scatter is large in each spiral sample.
Star formation properties of giant spiral samples and compact samples
are clearly different, implying that the star formation is regulated 
by the inner galaxy environment.

\item Only one KUG exhibits an explicit interaction feature, although half of 
our sample have some companions.
However, most of the spiral subset of our sample show distortions or 
isophotal twists.
This indicates that a weak encounter may have activated the star formation,
though we should note that their morphological distortion is not 
necessarily due to an external force.

\item The age and strength of star formation in individual 
samples has no relation to the existence of companions, implying that 
star formation may not be activated by the local environment.

\end{enumerate}

We should also note that there are no extremely peculiar galaxies in 
this region.
A detailed spatial distribution of KUGs and non-KUGs in the LUM filament 
should be investigated in order to further study the effects of environment.
A redshift survey to study the detailed spatial distribution of KUGs 
is now in progress, the results of which will appear in Takeuchi et al. (1998).

\acknowledgements

We first offer our thanks to the referee, Dr. J. W. Moody whose comments
greatly helped in improving the clarity and English presentation of this 
work.  We would like to thank Daisaku Nogami, Adel T. Roman, and 
Tadahiro Manmoto, as well as Kiso Observatory staff members for helping 
with observations. 
We also thank Masaru Hamabe for giving us the latest version of SPIRAL.
One of us (TTT) is grateful to Kouji Ohta, Takashi Ichikawa, 
Ichi Tanaka, Youichi Ohyama, and Shingo Nishiura for giving 
a great deal of useful suggestions and discussions. 
TTT also acknowledges the Research Fellowships
of the Japan Society for the Promotion of Science for Young
Scientists.
This research has made use of the NASA/IPAC Extragalactic Database (NED) 
which is operated by the Jet Propulsion Laboratory, California Institute of 
Technology, under contract with the National Aeronautics and Space 
Administration.
We also greatfully acknowledge the NASA's Astrophysics Data System Abstract 
Service (ADS).

\newpage
\begin{center}
{\bf FIGURE CAPTIONS}
\end{center}

Figure 1 -- The distribution of the fraction of Kiso Ultraviolet-excess 
galaxies (KUGs) $f_{\rm KUG}$ relative to the number density of CGCG 
galaxies $n$. 
Each symbol represents $f_{\rm KUG}$ in a $6^\circ \times
6^\circ$-area.
The confidence limits of 99.9 \%, 99.99 \%, and 99.999 \% are also shown.
The filled symbols, all outside the 99.99 \% envelop represent the KUG-rich regions. 
Eight filled symbols are plotted, but some of them (B and one of Ds, 
and C and one of Es) are overlapping with each other, and consequently 
the number of filled symbols appears as six.
The regions which have the same labels are adjacent to each other on
the sky.

Figure 2 -- All-sky projection map of the KUG-rich regions.
The small dots represent the distribution of CGCG galaxies in
KUG-survey 
regions, and the black hatches depict the KUG-rich regions.
The labels A, B, C, D, and E attached to the black-hatched regions
correspond to the labels in Fig. 1.

Figure 3 -- Contour maps in the $V$- and $I_{\rm C}$-bands of 
the objects in our sample.
They are shown from top to bottom in increasing order of right 
ascension.
Each contour interval is 0.5 magnitudes.
In this figure, the north is above, and the east to the right.
The FOV of the contour maps is $2.^\prime 5 \times 2.^\prime 5$.
A distance scale is presented in the upper-left of each galaxy
image.

Figure 4 -- The photometric partitioning of each galaxy image.

Figure 5 -- Color-color (C-C) diagrams of each square section 
of each galaxy shown in Fig. 4.
The data points labeled ``bulge'' and ``disk'' represent the color of bulge 
and disk sections of the object respectively.
The three broken lines on the C-C diagrams are the model starburst evolutionary
tracks superimposed on an older stellar population, given by 
Bica, Alloin, \& Schmidt (1990).

Figure 6 -- The $(V - R)^0_{\rm T}$ - $(V - I)^0_{\rm T}$ diagram, 
comparing the dereddened total colors of our KUGs in the Lynx-Ursa Major
filament with those covering the galaxy morphology 
sequence.
Open circles represent the mean color of each morphological type index $T$.
The shown indices are $-5$, $-4$, $-3$, $-2$, $-1$, 0, 1, 2, 3, 4, 
5, 6, $\{7\; {\rm and}\; 8\}$, 9, and 10, from Buta \& Williams 
(1995) and some are labeled to show the trend.
Error bars show the standard deviation of the color of galaxies in 
each morphological type.
Filled squares represent the dereddened color of our KUGs.
A vector showing the effect of the applied $A_V \sim 0.15$ reddening 
correctionis presented in the
top left of the diagram.  A typical photometric error of our data is
shown in the bottom right. 
The numbers beside the filled squares are the morphological type
indices of the KUGs. We give the symbol ``C'' to compact galaxies
instead of the $T$-index.
Our sample tends to deviate from the range of the $1\sigma$-strip of
the Hubble sequence.

\newpage

\begin{center}
TABLE 1 : {\sc KUG-Rich Regions}
\end{center}
\begin{center}
 \begin{tabular}{cccc}
  \hline
  \hline
 Label & R.A.& Dec.& KUG field No(s).$^a$ \\ \hline
 A & $9^{\rm h} \sim 10^{\rm h}$ & $42^\circ \sim 48^\circ $ & 0286, 0287, 0288 \\
 B & $\sim 12^{\rm h}$ & $36^\circ \sim 42^\circ $ & 0352, 0353, 0354 \\
 C & $7^{\rm h} \sim 8^{\rm h}$ & $30^\circ \sim 36^\circ $ & 0405, 0406 \\
 D & $\sim 13^{\rm h}$ & $18^\circ \sim 24^\circ $ & 0637 \\
  \hline
 \end{tabular}
\end{center}
\begin{enumerate}
\item[$a$] These are the original Kiso fields corresponding to the 
KUG-rich regions. 
Thus the locations and areas of them are slightly different from those shown
in Fig. 1.
\end{enumerate}

\newpage

\begin{center}
TABLE 2 : {\sc Lynx-Ursa Major Filament}
\end{center}
\begin{center}
 \begin{tabular}{lc}
  \hline
  \hline
  Location ($\alpha \,, \; \delta$) & $9^{\rm h} - 10^{\rm h}\, , \; 42^\circ 
- 48^\circ$ \\
  Projected Area$^a$ & $\sim 90$ deg$^2$ \\
  Velocity Range $cz$ & $2000 \; {\rm km\, s^{-1}} - 8000 \; {\rm km\, 
s^{-1}}$ \\
  Number of CGCGs $n$ & $\sim 120$ \\
  Number of KUGs $k$ & $\sim 70$ \\
  Fraction of KUGs $f_{\rm KUG}$ & $\sim 67 \; \%$ \\
  \hline
 \end{tabular}
\end{center}
\begin{enumerate}
\item[$a$] The filament extends east and west from the $6^\circ \times 6^\circ$
region A.
In the adjacent regions, KUG fractions are also significantly high at
The $\gnyoro 99.9$ \% confidence level.
\end{enumerate}

\newpage

\begin{center}
TABLE 3 : {\sc Selected KUGs and Their Basic Properties}
\end{center}
\begin{center}
 \begin{tabular}{cllrrrrrrccc}
  \hline
  \hline
  No. & KUG name & CGCG name  & \multicolumn{3}{c}{$\alpha_{1950.0}$} & \multicolumn{3}{c}{$\delta_{1950.0}$} & $cz^a$ & ${B^0_{\rm T}}^b$ & ${M_B}^c$ \\ \cline{4-12}
  ~ & ~ & ~ & $^{\rm h}$ & $^{\rm m}$ & $^{\rm s}$ & $^\circ$ & $'$ & $''$ & km$\,$s$^{-1}$ & mag & mag \\
  \hline 
 1  & 0908+451  & 0909.0+4510 & 9  & 8  & 55.7 & 45 &  9 & 39 & 2638 & 12.14 & $-$19.96 \\
 2  & 0908+468  & 0908.4+4651 & 9  & 8  & 18.0 & 46 & 50 & 42 & 4287 & 14.50 & $-$18.66 \\
 3  & 0911+471  & 0911.6+4707 & 9  & 11 & 34.5 & 47 &  6 & 38 & 4242 & 14.20 & $-$18.94 \\
 4  & 0919+474  & 0919.0+4727 & 9  & 19 &  5.0 & 47 & 27 & 28 & 9121 & 15.77 & $-$19.03 \\
 5  & 0924+448  & 0925.0+4453 & 9  & 24 & 55.1 & 44 & 52 & 56 & 7717 & 14.18 & $-$20.26 \\
 6  & 0944+468  & 0944.2+4651 & 9  & 44 &  7.1 & 46 & 50 & 31 & 4708 & 14.74 & $-$18.62 \\
 7  & 0945+443  & 0945.5+4418 & 9  & 45 & 34.2 & 44 & 18 & 49 & 4786 & 13.30 & $-$20.10 \\
 8  & 0947+445A & 0947.0+4432 & 9  & 47 &  1.8 & 44 & 31 & 43 & 4666 & 14.52 & $-$18.83 \\
 9  & 0953+466  & 0953.6+4642 & 9  & 53 & 32.6 & 46 & 41 & 57 & 4681 & 15.15 & $-$18.20 \\
 10 & 1007+461  & 1007.8+4612 & 10 & 07 & 48.4 & 46 & 11 & 49 & 7291 & 14.40 & $-$19.91 \\
 11 & 1016+467  & 1016.0+4643 & 10 & 16 &  0.7 & 46 & 42 & 21 & 9169 & 14.10 & $-$20.71 \\
  \hline
 \end{tabular}
\end{center}
\begin{enumerate}
\item[$a$] Galactocentric velocities based on the value taken from the NED. 
\item[$b$] $B^0_{\rm T}$-magnitudes from RC3.
\item[$c$] $M_B$ for $h = 1.0$. 
\end{enumerate}

\newpage

\begin{center}
TABLE 4 : {\sc Observation Log}
\end{center}
\begin{center}
 \begin{tabular}{lll}
  \hline \hline
   KUG name & Date & Frame ID$^a$ \\ \hline
   0908+451  & 1996 Nov 25 & 36782 $\sim$ 36790 \\
   0908+468  & 1996 Nov 28 & 37087 $\sim$ 37095 \\
   0911+471  & 1996 Nov 28 & 37099 $\sim$ 37107 \\
   0919+474  & 1996 Nov 27 & 36971 $\sim$ 36979 \\
   0924+448  & 1996 Nov 28 & 37061 $\sim$ 37069 \\
   0944+468  & 1996 May 27 & 31375 $\sim$ 31383 \\
   0945+443  & 1996 May 27 & 31387 $\sim$ 31395 \\
   0947+445A & 1996 Jun 1  & 31622 $\sim$ 31630 \\
   0953+466  & 1996 Nov 27 & 36959 $\sim$ 36967 \\
   1007+461  & 1996 Nov 28 & 37111 $\sim$ 37119 \\
   1016+467  & 1996 Jun 1  & 31634 $\sim$ 31642 \\
  \hline
 \end{tabular}
\end{center}
\begin{enumerate}
\item[$a$] Frame ID numbers of the objects used in our paper.
The actual name is of the form kcc36782, and here we omit the prefix
``kcc''.
Quick looks of the raw data of our observations are available on the World
Wide Web through the Mitaka-Okayama-Kiso data Archival system (MOKA) at 
http://www.moka.nao.ac.jp.
MOKA is operated by Astronomical Data Analysis Center, Okayama 
Astrophysical Observatory (National Astronomical Observatory of Japan) and 
Kiso Observatory (University of Tokyo) in cooperation wich the Japan 
Association Information Processing in Astronomy (Horaguchi et al. 1994; 
Takata et al. 1995).
\end{enumerate}

\newpage

\begin{center}
TABLE 5 : {\sc Total Magnitudes by Growth-curve Fitting}
\end{center}
\begin{center}
\begin{tabular}{lccccccc}
\hline \hline
KUG name & $V$ & $R_{\rm C}$ & $I_{\rm C}$ & ~ & $M_V$ & $M_R$ & $M_I$ \\
\cline{2-4} \cline{6-8}
~ & ~ & mag & ~ & ~ & \multicolumn{3}{c}{$+\; 5 \log h$ mag} \\
\hline 
0908+451 & 11.54 & 11.01 & 10.63 & ~ & $-$20.65 & $-$21.16 & $-$21.51 \\
0908+468 & 13.97 & 13.56 & 13.22 & ~ & $-$19.28 & $-$19.67 & $-$19.98 \\
0911+471 & 13.47 & 12.99 & 12.56 & ~ & $-$19.76 & $-$20.22 & $-$20.62 \\
0919+474 & 15.28 & 14.88 & 14.63 & ~ & $-$19.61 & $-$19.99 & $-$20.21 \\
0924+448 & 13.49 & 13.02 & 12.58 & ~ & $-$21.04 & $-$21.49 & $-$21.90 \\
0944+468 & 14.36 & 13.96 & 13.51 & ~ & $-$19.09 & $-$19.47 & $-$19.89 \\
0945+443 & 12.61 & 12.17 & 11.72 & ~ & $-$20.88 & $-$21.30 & $-$21.72 \\
0947+445A & 13.88 & 13.39 & 12.91 & ~ & $-$19.56 & $-$20.03 & $-$20.48 \\
0953+466 & 14.59 & 14.10 & 13.68 & ~ & $-$18.85 & $-$19.32 & $-$19.71 \\
1007+461 & 13.73 & 13.32 & 12.98 & ~ & $-$20.67 & $-$21.06 & $-$21.37 \\
1016+467 & 13.68 & 13.28 & 12.90 & ~ & $-$21.22 & $-$21.60 & $-$21.95 \\
\hline
\end{tabular}
\end{center}

\newpage

\begin{center}
TABLE 6 : {\sc Comparison of Our Result with Literature}
\end{center} 
\begin{center}
 \begin{tabular}{cccc}
  \hline \hline
   Object & Band & Our Photometry & Huchra 1977 \\ \hline
   KUG0953+466 & $V$ & $14.96 \pm 0.03$ & $14.92 \pm 0.05$ \\
   ~ & $R_{\rm C}$ & $14.46 \pm 0.03$ & $14.41 \pm 0.03$ \\
  \hline
 \end{tabular}
\end{center}

\newpage

\begin{center}
TABLE 7 : {\sc Integrated Colors}
\end{center}
\begin{center} 
 \begin{tabular}{lccc}
  \hline \hline
  KUG name & $(V - R_{\rm C})_{\rm T}$ & $(V - I_{\rm C})_{\rm T}$ & $(R_{\rm C} - I_{\rm C})_{\rm T}$ \\ \hline
   0908+451  & 0.51 & 0.86 & 0.35 \\
   0908+468  & 0.39 & 0.70 & 0.31 \\
   0911+471  & 0.46 & 0.86 & 0.40 \\
   0919+474  & 0.38 & 0.60 & 0.22 \\
   0924+448  & 0.45 & 0.85 & 0.41 \\
   0944+468  & 0.38 & 0.80 & 0.42 \\
   0945+443  & 0.42 & 0.84 & 0.42 \\
   0947+445A & 0.48 & 0.92 & 0.45 \\
   0953+466  & 0.47 & 0.85 & 0.39 \\
   1007+461  & 0.39 & 0.71 & 0.31 \\
   1016+467  & 0.39 & 0.73 & 0.35 \\
  \hline
 \end{tabular}
\end{center}

\newpage

\begin{center}
TABLE 8: {\sc Summary on Morphology and Star Formation Characteristics}
\end{center}
{\scriptsize
\begin{center} 
 \begin{tabular}{llcllllll}
  \hline \hline
  KUG name &\multicolumn{2}{c}{Type}& SF$^a$ & Age$^b$ & Burst & \multicolumn{3}{c}{Notes} \\ \cline{2-3}\cline{7-9}
  ~ & $T$ & KUG & ~& [yr] & [\%] & Appearance & Distortion & Environment\\ \hline
   0908+451  & 5     & Sk  & Disk  & $10^{7 - 8}$ & $1-10$ & Knotty & Yes & Isolated\\
   0908+468  & C$^c$ & C   & Whole & $10^7$       & $10$   & Featureless & $-$ & Isolated \\
   0911+471  & 5$^d$ & Sk  & Disk  & $10^8$       & $10$   & Knotty & Yes & Isolated \\
   0919+474  & C$^c$ & C   & Whole & $10^{6 - 7}$ & $10$   & Featureless & $-$& In a group \\
   0924+448  & 5     & Sk  & Disk  & $10^8$       & $1-10$ & Knotty & Yes (triple-armed) & Isolated \\
   0944+468  & 6     & Sk  & Disk  & $10^{7 - 8}$ & $1-10$ & Knotty & Yes & Isolated \\
   0945+443  & 5     & Sk  & Disk  & $10^8$       & $10$   & Knotty & $-$ & In a group \\
   0947+445A & 4$^d$ & Sp: & Disk  & $10^{8 - 9}$ & $1-10$ & Smooth & $-$ (four-armed) & In a group \\
   0953+466  & C$^c$ & C   & Whole & $10^{6 - 7}$ & $\lnyoro 1$ &Featureless & $-$ & Isolated \\
   1007+461  & 5$^d$ & Sk  & Disk  & $10^{7 - 8}$ & $10$   & Smooth & $-$ & Companion \\
   1016+467  & 4     & Pi: & Disk  & $10^{7 - 8}$ & $10$   &  $-$ & Yes (tidal feature) & Inteacting \\
  \hline
 \end{tabular}
\end{center}
}
\begin{enumerate}
\item[$a$] Location where active star formation occurs on the galaxy.
\item[$b$] Burst-to-galaxy mass ratio based on Bica et al. (1990).
\item[$c$] Not given in RC3, and has featureless compact morphology.
\item[$d$] Not given in RC3. Assigned by us. 
\end{enumerate}
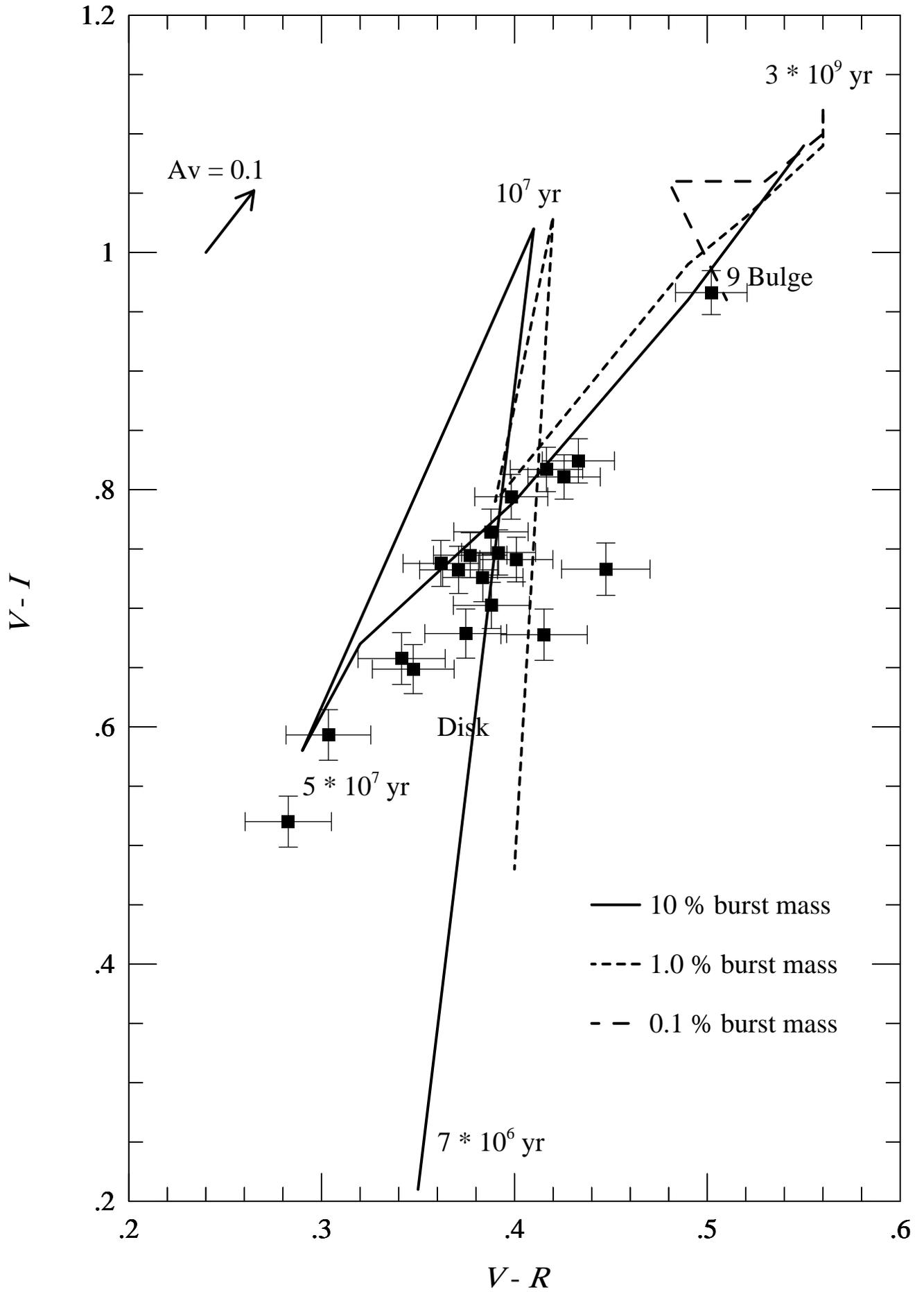
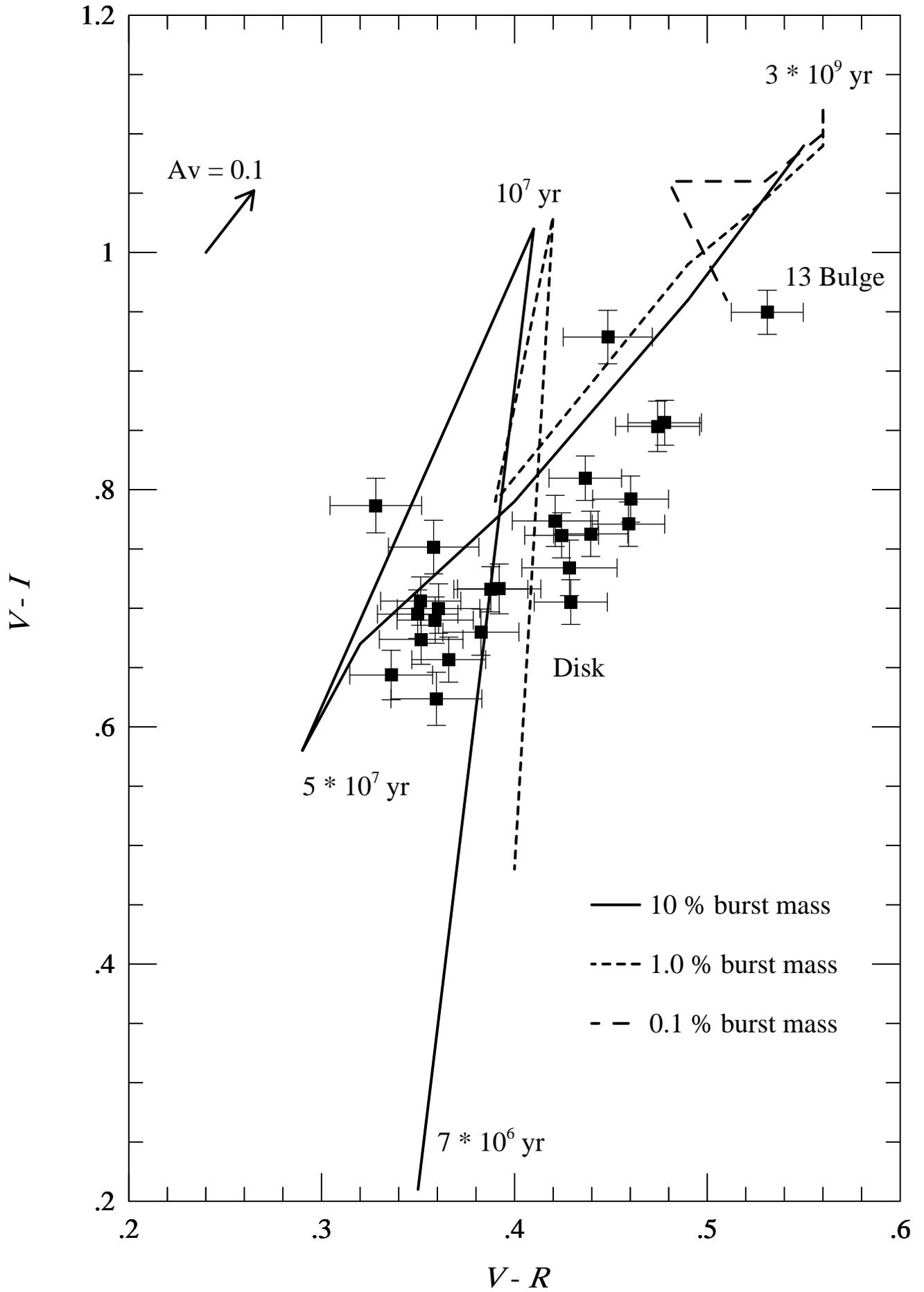
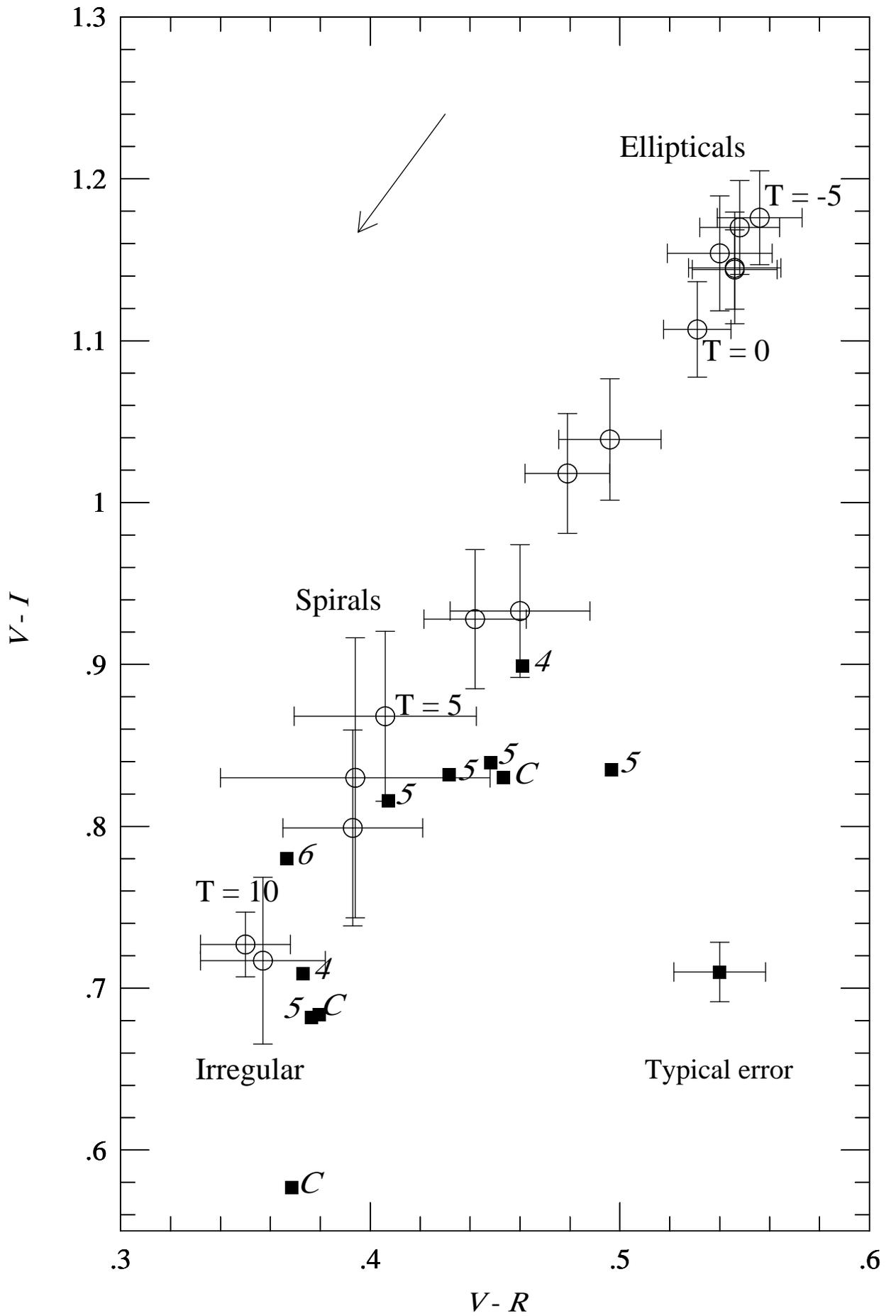
\end{document}